\documentclass[11pt,a4paper,reqno]{amsart}
\usepackage{multirow}
\usepackage[centertags]{amsmath}
\usepackage{amsfonts}
\usepackage{amssymb}
\usepackage{amsthm}
\usepackage{newlfont}
\usepackage{mathtools}
\usepackage{a4}
\usepackage{enumerate}
\usepackage[pdftex]{graphicx,color}
\usepackage{diagbox}

\newcommand{\ben}{\begin{equation}}
\newcommand{\een}{\end{equation}}
\newcommand{\bens}{\begin{equation*}}
\newcommand{\eens}{\end{equation*}}
\newcommand{\bal}{\begin{align}}
\newcommand{\eal}{\end{align}}
\newcommand{\nn}{\nonumber\\ }

\newcommand{\infinity}{{\infty}}

\newcommand{\ad}{{\mathrm{ad}}}

\newcommand{\hfb}{{\hfill\break}}

\input amssym.def
\def\Z{{\mathbb Z}}   \def\N{{\mathbb N}}
   
\begin{document}
\baselineskip=18pt
\parskip=8pt

\title[Adjoint affine fusion]
{Adjoint affine fusion and tadpoles}

\author[A. Urichuk]{Andrew Urichuk$^{1}$}
\author[M.A. Walton]{Mark A. Walton$^{1, 2}$}

\date{\today}
\vskip1cm
\begin{abstract}
\large\baselineskip=16pt 
${}$ \\ We study affine fusion with the adjoint representation.   For simple Lie algebras, elementary and universal formulas determine the decomposition of a tensor product of an integrable highest-weight representation with the adjoint representation.  Using the (refined) affine depth rule, we prove that equally striking results apply to adjoint affine fusion.  For diagonal fusion, a coefficient equals the number of nonzero Dynkin labels of the relevant affine highest weight, minus 1.  A nice lattice-polytope interpretation follows, and allows the straightforward calculation of the genus-1 1-point adjoint Verlinde dimension, the adjoint affine fusion tadpole.  Explicit formulas, (piecewise) polynomial in the level, are written for the adjoint tadpoles of all classical Lie algebras.  We show that off-diagonal adjoint affine fusion is obtained from the corresponding tensor product by simply dropping non-dominant representations. \\ 
\end{abstract}

\maketitle
\noindent
$^1$ Physics \& Astronomy Department, University of Lethbridge, Lethbridge, Alberta\  T1K 3M4, Canada\\   
$^2$ International School for Advanced Studies (SISSA), via Bonomea 265, 34136 Trieste, Italy\\

\noindent
\textit{E-mail:} andrew.urichuk@uleth.ca, walton@uleth.ca

\vfill\eject

\section{Introduction} 
Affine fusion is realized in Wess-Zumino-Witten    \cite{Witten84, GW86} conformal field theories    \cite{DMS97}, and  integrable models  
\cite{PS90, GN91,KS10, Korff11, MWphase},  
for examples.  While  associated with nontwisted affine Kac-Moody algebras at fixed integer level, it can be regarded as a nice truncation of the tensor product (decomposition)  of simple, complex Lie algebras, controlled by the affine level  \cite{CMW91, KMSWconf92, KMSW93, MW94}. Mathematically, affine fusion also describes the tensor products for quantum groups   \cite{PS90, FGP90} and Hecke algebras at roots of unity   \cite{GW90}, and is relevant to the study of knot invariants \cite{WittenKnots, Turaev} and quantum cohomology   \cite{WittenQCoh, Fulton04, MS12}. 

Much is now known about affine fusion.  For example, the Verlinde formula    \cite{Verlinde88} combined with the Kac-Peterson modular $S$ matrix    \cite{KP84} leads to an algorithm    \cite{Kac90, MW90, FGP90} computing the affine fusion coefficients as  alternating sums over (finite sets of short) elements of  the affine Weyl group.  However, a general and combinatorial algorithm without cancellations, a long-sought\footnote{\,See    \cite{MW94, MW98, Tud00} , e.g. for some older attempts, and    \cite{MS12} and references therein for recent progress, as well as mathematical motivation.} `Littlewood-Richardson rule' for fusion,  has not yet  been found.      

The (original) Gepner-Witten depth rule    \cite{GW86} for affine fusion involves no cancellations, but is difficult to implement \cite{KMSW93,KMSWconf92}. It is relevant to the hunt for a Littlewood-Richardson rule for fusion, however. A  formula similar to the depth rule exists for the tensor product coefficients   \cite{Zhelobenko, PRV67, Kostant} of the horizontal, finite-dimensional subalgebra, and it leads to a proof of the original Littlewood-Richardson algorithm for $A_r\cong s\ell(r+1)$.  The similarity with this `finite depth rule' was noticed in   \cite{KMSW93, KMSWconf92}, where the refined, affine depth rule was conjectured.  The affine formula was finally proved in   \cite{FF08}. 

Here we will demonstrate the usefulness of the affine depth rule directly,  instead of pursuing a Littlewood-Richardson rule.\footnote{\,A similar approach was taken in   \cite{MW94}.}   We will restrict attention to a specific, important  class of representations:  the adjoint representations of the complex  simple Lie algebras.  Such a specialization is recommended by the simplicity and significance of affine fusion with the fundamental representations of the $A_r$ algebras.  The so-called Pieri rule is a simple truncation of that for $A_r$ tensor products,  leads to a nice Giambelli formula and Schubert calculus for affine fusion   \cite{Gep91}, and characterizes $A_r$ fusion  \cite{GW90}.  

We use the affine depth rule to write remarkably simple formulas for adjoint fusion -- see equations  (\ref{DiagAdjFusDLs}, \ref{NcOff}).  They are universal (uniform for all simple Lie algebras) and valid for all ranks and algebras. 

For diagonal fusion, the result is particularly elementary, mimicking the analogous formula for diagonal adjoint tensor product coefficients \cite{Kostant, BZ89}, and giving the fusion coefficient a pretty lattice polytope interpretation.  As a consequence,  the adjoint affine fusion tadpole\footnote{\,This term was introduced for Feynman diagrams in   \cite{CG64}.}  (the genus-1 1-point Verlinde fusion dimension)  can be computed in a straightforward way.  The fusion tadpole is interesting as the simplest non-trivial example of a positive-genus fusion (or Verlinde) dimension.  

Formulas that are piecewise polynomial in the level can be written for the adjoint fusion tadpoles of any simple Lie algebra, by considering them as integer-weighted sums on the lattice polytope defined by the set of possible highest weights.  Explicit expressions are given for adjoint tadpoles of all classical Lie algebras, and for the exceptional algebra $E_6$.\footnote{\,Affine fusion tadpoles were studied in   \cite{Uri15}, where some  explicit formulas of this kind were written.}  The last result demonstrates that there is no obstruction to writing the formulas for the remaining exceptional algebras.  The number of distinct pieces for the piecewise polynomial formulas can be large, however.

In the next section, we will  introduce our notation, and write the refined depth rule and its analogue for simple Lie algebras (we  call these the affine and finite depth rules, respectively).  In section 3, the affine depth rule will be applied to adjoint fusion.  Simple, general formulas will be written separately for the diagonal and off-diagonal fusion coefficients.  Section 4 applies the diagonal result to the calculation of affine fusion tadpoles. It includes a table of explicit tadpole formulas. Section 5 is a short conclusion.      

\section{Finite and Affine Depth Rules}

Let $X_r$ denote a simple Lie algebra of  rank $r$.  The corresponding nontwisted affine Lie algebra is usually written $X_r^{(1)}$; here $X_{r,k}$ will stand for the affine algebra at fixed level $k$.    $X_r$ is embedded in $X_{r,k}$ as its horizontal subalgebra.  Put $I:=\{1,2,\ldots,r\}$ and $\hat I:=\{0,1,2,\ldots,r\}$.  

Let $F \coloneqq \{ \Lambda^i |  i\in I\}$ denote the set of fundamental weights of $X_r$, and $\hat F\coloneqq \{ \hat\Lambda^i |  i\in \hat I \}$ that of $X_{r,k}$. The weight lattice of $X_r$  is $P:=\Z F$, and that of $X_{r,k}$ is $\hat P :=\Z \hat F$.  $(\cdot, \cdot)$ will denote the inner product of weights. A weight $\lambda\in P$ ($\hat\lambda\in \hat P$) has the expansion $\lambda=\sum_{i\in I} \lambda_i\Lambda^i$  ($\hat\lambda=\sum_{i\in \hat I} \lambda_i \hat\Lambda^i$); the $\lambda_{i\in I}$ ($\lambda_{i\in\hat I}$) are the finite (affine) Dynkin labels of $\lambda$ ($\hat\lambda$).  

Let $\Delta$ be the set of roots of $X_r$, $\Delta_+$  ($\Delta_-$) the positive (negative) roots, and $S := \{ \alpha_i | i\in I \}$ the set of simple roots. $\theta$ will stand for the highest root of $X_r$. With $\alpha^\vee_i := 2\alpha_i/(\alpha_i, \alpha_i)$ a simple co-root,  the decomposition $\theta = \sum_{i\in I} m^\vee_i\alpha^\vee_i$ defines the co-marks $m^\vee_i\in \N$, $i\in I$. 
 
The primitive reflection $r_i$ in weight space reflects across a (hyper-)plane through the origin with normal  direction determined by the simple root $\alpha_i$, $i\in I$. Therefore, for any weight $\lambda\in P$,  $r_i\lambda = \lambda - (\alpha_i^\vee, \lambda)\alpha_i = \lambda - \lambda_i\alpha_i$.  The Weyl group $W$ of $X_r$ is a finite group generated by the primitive reflections:  $W = \langle\, r_i \,|\, i\in I  \,\rangle$. 

Define the Weyl vector $\rho:=\sum_{i\in I} \Lambda^i = \frac 1 2 \sum_{\alpha\in \Delta_+} \alpha$. The shifted action of primitive reflections (and all elements of the Weyl group $W$) is also useful:  $r_i.\lambda := r_i(\lambda+\rho)-\rho = \lambda - (\lambda_i+1)\alpha_i$. 

An irreducible integrable highest-weight representation  of $X_r$, of highest weight $\lambda$, will be written $L(\lambda)$. A highest weight $\lambda = \sum_{i\in I} \lambda_i \Lambda^i$ has non-negative integer Dynkin labels, and so must lie in the cone  $P_+ := \{ \lambda\in P | \lambda_j \in \N_0, j\in I \}$.  The analogous affine definition is $\hat P_+ := \{ \hat\lambda= \sum_{j\in\hat I} \lambda_j\hat\Lambda^j \in \hat P \,|\, \lambda_j \in \N_0, j\in \hat I \}$. 

The highest weight $\hat\lambda = \sum_{j\in\hat I} \lambda_j \hat\Lambda^j$ of an irreducible integrable representation of the affine Lie algebra $X_{r,k}$ has non-negative integer  Dynkin labels also satisfying $\lambda_0 + \sum_{i\in I} m^\vee_i \lambda_i = k$.  Putting $m^\vee_0 := 1$, the affine highest weights belong to the set $\hat P_+^k := \{ \lambda\in \hat P_+ | \sum_{j\in \hat I} m^\vee_j \lambda_j = k \}$.  This set of affine dominant weights projects to the following set of horizontal $X_r$ dominant weights: $P_+^k := \{  \lambda\in P_+ | \lambda = \sum_{i\in I} \lambda_i \Lambda^i \, |\, \sum_{i\in I} m^\vee_i\lambda_i = (\theta, \lambda) \leq k \}$.  $P_+^k$ is a projection of  the set $\hat P_+^k$ of affine dominant weights to $P_+$, forming a lattice polytope, a truncation of the cone $P_+$ of dominant weights, a fragment of the weight lattice $P$ of $X_r$.  

We will use the Cartan-Weyl basis of the Lie algebra $X_r$ (see Ch. 6 of   \cite{FuchsSchweigert97}, e.g.).  Generators $H_\alpha$ and $E_\alpha$ are associated to each root $\alpha\in \Delta$ and the commutation relations are
\begin{align}
[H_\alpha, H_\beta ] \, =\, 0\ ,\ \ &\ \ [E_\alpha, E_{-\alpha}]\ =\  H_\alpha\ , \label{HHEE}\\   
[H_\alpha, E_\beta ]\,  =&\,  (\alpha^\vee, \beta)\, E_\beta\ \ , \label{HE}\\ 
[E_\alpha, E_\beta ]\ =\ &e_{\alpha, \beta}\, E_{\alpha+\beta}\ ,\ \ \ {\text {for}}\ \alpha\not=\beta\ ;\label{EE} 
\end{align}
for all $\alpha, \beta\in \Delta$.  In (\ref{EE}), $e_{\alpha, \beta}\not=0$ whenever $\alpha+\beta\in \Delta$. 

A highest-weight representation $L(\lambda)$ of $X_r$ contains subspaces $L_\mu(\lambda)$ of fixed weight $\mu$.  Put $H_i := H_{\alpha_i}$ for $\alpha_i\in S$, and write $H:= \sum_{j\in I} \Lambda^j H_j$.  Then  
\ben L_\mu(\lambda)\ :=\ \left\{\,  v\in L(\lambda) \ \vert\  H\, v = \mu\, v  \,\right\}\ . \label{wtsub}\een 

Consider now the tensor product decomposition of 2 highest weight representations of $X_r$:
\ben L(\lambda)\otimes L(\mu)\ \hookrightarrow\ \bigoplus_{\nu\in P_+} c_{\lambda,\mu}^\nu\ L(\nu)\ \ .\label{tp} \een  
For the algebra $A_r$, the tensor product multiplicities $c_{\lambda, \mu}^\nu$ are the Littlewood-Richardson coefficients. For all simple Lie algebras $X_r$, they satisfy the universal formula   \cite{Zhelobenko, PRV67, Kostant}
\ben  c_{\lambda,\mu}^\nu\ =\ \dim\left\{ v\in L_{\nu-\mu}(\lambda)\ \big\vert\  \left( E_{-\alpha_i} \right)^{\nu_i+1}\,v=0\, ,\ \forall i\in I \,  \right\}\ ;\label{fdepthf}\een
and the equivalent
\ben  c_{\lambda,\mu}^\nu\ =\ \dim\left\{ v\in L_{\nu-\mu}(\lambda)\ \big\vert\  \left( E_{+\alpha_i} \right)^{\mu_i+1}\,v=0\, ,\ \forall i\in I \,  \right\}\ .\label{fdepthe}\een 
These constitute the finite depth rule.\footnote{\, This was called the PRV Theorem in \cite{Stem}.} 

Replacing the tensor product $\otimes$ with the fusion product $\otimes_k$  at level $k$ truncates  (\ref{tp}) to
\ben L(\lambda)\otimes_k L(\mu)\ \hookrightarrow\ \bigoplus_{\nu\in P_+} {}^{(k)}N_{\lambda,\mu}^\nu\ \, L(\nu)\ \ . \label{fusk} \een  The fusion multiplicities (coefficients) indeed truncate the result  (\ref{tp}), since ${}^{(k)}N_{\lambda,\mu}^\nu \leq c_{\lambda,\mu}^\nu$.  

The refined depth rule   \cite{KMSW93, KMSWconf92, FF08} provides the affine analogs
\ben  {}^{(k)}N_{\lambda,\mu}^\nu\ =\ \dim\left\{ v\in L_{\nu-\mu}(\lambda)\ \big\vert\  \left( E_{-\alpha_i} \right)^{\nu_i+1}\,v=0\, ,\ \forall i\in I \ ;\  \left( E_{+\theta} \right)^{k-(\theta,\nu)+1}\,v=0\    \right\}\ ;\label{adepthf}\een
and 
\ben  {}^{(k)}N_{\lambda,\mu}^\nu\ =\ \dim\left\{ v\in L_{\nu-\mu}(\lambda)\ \big\vert\  \left( E_{+\alpha_i} \right)^{\mu_i+1}\,v=0\, ,\ \forall i\in  I \, ;\, \ \left( E_{-\theta} \right)^{k-(\theta,\mu)+1}\,v=0\,   \right\}\ .\label{adepthe}\een 
Notice that  for large level $k$, the additional conditions in (\ref{adepthf}, \ref{adepthe}) beyond the finite ones  are trivially satisfied, so that  the finite depth rule (\ref{fdepthf},\ref{fdepthe}) is recovered. 

The similarity between the finite (horizontal) and affine depth rules becomes clear when one notes that a level-$k$ affine dominant weight $\hat\nu$ has Dynkin label $\nu_0=k-(\theta,\nu)$, and the affine generators $\hat E_{\pm\alpha_0}$ project to the horizontal simple Lie algebra generators $E_{\mp\theta}$.  To make this explicit, let us put\footnote{\,It will also be useful later to define the reflection $r_0$ by the action $r_0\lambda :=  \lambda + \lambda_0\theta$ on $\lambda\in P$.} 
\begin{align} 
E_{\pm\alpha_0}\ :=\ E_{\mp\theta}\ \ .\label{Ezero}
\end{align}
Then the affine depth rule takes the form
\ben  {}^{(k)}N_{\lambda,\mu}^\nu\ =\ \dim\left\{ v\in L_{\nu-\mu}(\lambda)\ \big\vert\  \left( E_{-\alpha_i} \right)^{\nu_i+1}\,v=0\, ,\ \forall i\in \hat I \    \right\}\ ;\label{adepthfsymm}\een
and 
\ben  {}^{(k)}N_{\lambda,\mu}^\nu\ =\ \dim\left\{ v\in L_{\nu-\mu}(\lambda)\ \big\vert\  \left( E_{+\alpha_i} \right)^{\mu_i+1}\,v=0\, ,\ \forall i\in  \hat I \   \right\}\ .\label{adepthesymm}\een 
The mixed finite-affine nature of affine fusion is apparent in these formulas.

\section{Adjoint Affine Fusion}

Let us now apply the affine depth rule to calculate adjoint fusion.  For an arbitary representation $L(\lambda)$ of $X_r$, $\lambda\in P_+$,  the refined affine depth rule is not straightforward to use \cite{KMSW93, KMSWconf92}, but the adjoint representation is special.  $L(\theta)$ has a simple weight system, consisting of the roots $\Delta$, with multiplicity 1,  and $r$ copies of the 0 weight.  Most importantly, however, the adjoint representation is defined by $x\, \rightarrow \ad[x]$, where $x\in X_r$, and  $ \ad[x]\, y\, =\, [x, y]$. The commutation relations of the Cartan-Weyl basis (\ref{HHEE}-\ref{EE}) then allow us to proceed. 

We will find simple results that reduce to similar formulas for the finite case when the level $k\to\infinity$.

\subsection{Diagonal Adjoint Affine Fusion} 

First consider diagonal adjoint affine fusion, with coefficients ${}^{(k)}N_{\theta,\mu}^\mu$. Parametrize the vector $v\in L_{0}(\theta)$ as \ben v\ =\ \sum_{j\in I}\, v_j\, H_{\alpha_j}\ \ ,\label{vparam}\een Then 
\ben   {}^{(k)}N_{\theta,\mu}^\mu\ =\ \dim\left\{ \sum_{j\in I} c_j\, H_{\alpha_j}\  \bigg\vert\   \ad[E_{\alpha_i}]^{\,\mu_i+1}\,\sum_{j\in I} c_j\, H_{\alpha_j}\,=0\, ,\ \forall i\in  \hat I \  \right\}\ .\label{DepthThetaDiag}\een 

The commutation relation (\ref{HE}) shows that  $\ad[E_{\alpha_i}]^{\, 2} H_{\alpha_j} = 0$, so the $i$-th constraint in (\ref{DepthThetaDiag}) is satisfied if $\mu_i >0$.  

Putting $\tilde v := \sum_{j\in I} v_j \alpha_j^\vee$, we find that for any $i\in \hat I$, 
\ben  \mu_i\ =\ 0\ \ \rightarrow\ \ (\alpha_i, \tilde v)\ =\ 0\ \ . \label{vtilde}\een  Since $\alpha_0=-\theta$, the inner products $ (\alpha_i, \tilde v)$ are not all independent.  But those for any proper subset of $\{ \alpha_i\ \vert\ i\in\hat I \}$ are.\footnote{\,One way to see this is to consider the extended Dynkin diagram (isomorphic to the affine Dynkin diagram) with $r+1$ nodes labelled by the $\alpha_j$, $j\in\hat I$.  Deleting any number of nodes of the extended diagram leaves the Dynkin diagram of a semi-simple subalgebra of $X_r$, with nodes labelled by its simple roots, all independent.}  Since the level $k$ must be nonzero for the adjoint representation to appear ($\theta\in P_+^k\, \rightarrow\, k\geq 2$), at least one `affine' Dynkin label $\mu_i\not=0\ (i\in\hat I)$. Therefore, the constraints imposed by $\mu_i=0$, $i\in \hat I$, will be independent. 

Consequently, the dimension of (\ref{DepthThetaDiag}) is reduced by 1 for every vanishing affine Dynkin label $\mu_i,\ i\in\hat I$. We therefore have 
\ben   {}^{(k)}N_{\theta,\mu}^\mu\ =\ \dim\left\{\, \mu_j\not=0,\  j\in\hat I  \,\right\} \, -\, 1\ \ ;\label{DiagAdjFusDLs}\een a diagonal adjoint affine fusion coefficient equals the number of nonzero affine Dynkin labels $\mu_i$, $i\in \hat I$, minus 1. 

If the level $k$ is large, then the Dynkin label $\mu_0\not=0$.  Then the classic result \cite{Kostant, BZ89} for tensor product coefficients, 
\ben   c_{\theta,\mu}^\mu\ =\ \dim\left\{\, \mu_j\not=0,\  j\in I  \,\right\} \ \  \label{DiagAdjTPDLs}\een is recovered. A diagonal adjoint tensor product coefficient equals the number of nonzero (finite) Dynkin labels $\mu_i$, $i\in I$.

\subsection{Off-Diagonal Adjoint Affine Fusion} 

For completeness, let us also consider off-diagonal adjoint fusion.  By the affine depth rule (\ref{adepthfsymm}, \ref{adepthesymm}), a nonzero  coefficient ${}^{(k)}N^\nu_{\theta, \mu\not=\nu}$ must have 
$\nu-\mu \in \Delta$.  Adjoint affine fusion can only be a little off-diagonal, by a root. 

The space $L_\beta(\theta)$ of weight $\beta$ in the adjoint representation $L(\theta)$ is 1-dimensional, with single basis element $E_\beta$.  Eqn.  (\ref{adepthesymm}) gives
\ben    {}^{(k)}N_{\theta, \mu}^{\nu\not=\mu}\ =\ 
  \begin{cases}
      1\, , & \text{if}\  \ \nu-\mu\in\Delta\ \ \,\text{and}\ \ \,\ad[E_{\alpha_i}]^{\mu_i+1} E_{\nu-\mu}  = 0\, ,\  \forall\, i\in\hat I\  ; \\
      0\, ,  &  \text{otherwise}\, . 
  \end{cases}  
\een 
First suppose that $\nu-\mu\not= -\alpha_i$; then the commutation relation (\ref{EE}) allows us to analyze the condition $\ad[E_{\alpha_i}]^{\mu_i+1} E_{\nu-\mu} \, =\, 0$  easily.  Since the structure constants $e_{\alpha, \beta}$ in (\ref{EE}) are all non-vanishing for $\alpha,  \beta, \alpha+\beta\in\Delta$, it  implies that $\nu-\mu+(\mu_i+1)\alpha_i\not\in\Delta$, or 
\ben \nu-\mu+(\mu_i+1)\alpha_i\ =\ \nu-r_i.\mu\ \not\in\ \Delta\ .     \label{numui}\een
Note that since $-\Delta = \Delta$ and $r_i\Delta = \Delta$,  the condition $\nu-r_i.\mu \not\in \Delta$ can be written alternatively as $r_i.\nu-\mu \not\in \Delta$, $\mu-r_i.\nu \not\in \Delta$, or $r_i.\mu-\nu \not\in \Delta$, for any $i\in \hat I$. 

Now suppose that $\nu-\mu=-\alpha_i$. Then $\nu-r_i.\nu=0 \not\in\Delta$ is possible,  iff $\mu_i=0$. In this case,  the $i$-th condition is not satisfied: 
\ben  
\ad[E_{\alpha_i}]^{\mu_i+1} E_{-\alpha_i} \, =\,  H_{\alpha_i}\, \not=\, 0\, .  
\een
If $\mu_i=1$, then 
\ben  
\ad[E_{\alpha_i}]^{\mu_i+1} E_{-\alpha_i} \, =\, \ad[E_{\alpha_i}] H_{\alpha_i}\, =\, -(\alpha_i^\vee, \alpha_i) E_{\alpha_i} \not=\, 0\, ,  
\een
but then  $\nu-r_i.\mu=\alpha_i\in\Delta$.  For $\mu_i>1$, the condition is satisfied, and so is $\nu-r_i.\mu\not\in \Delta$.  

We therefore find  
\ben    {}^{(k)}N_{\theta, \mu}^{\nu\not=\mu}\ =\ 
  \begin{cases}
      1\, , & \text{if}\  \ \nu-\mu\in\Delta\ \ \,\text{and}\ \ \,\nu-r_i.\mu \not\in \Delta\cup\{0\}\, ,\ \,   \forall\, i\in\hat I \ ; \\
      0\, ,  &  \text{otherwise}\, . 
  \end{cases}  
\label{OffDiagF}
\een 
The result (\ref{OffDiagF}) is simple in that 1) it involves no cancellations, and 
2) no knowledge of non-trivial weight multiplicities is required. One only needs the set of roots $\Delta$. 

As an example, Table~\ref{G2table}  displays the conditions on the Dynkin labels of a weight $\mu$ from (\ref{OffDiagF}) such that  ${}^{(k)}N_{\theta,\mu}^\nu=1$.  Note that most of the constraints from (\ref{OffDiagF}) are trivial, in the sense that if the weights involved are in $P_+^k$, they are satisfied automatically. Only 4 of the roots of $G_2$ give rise to non-trivial conditions, $\alpha_1+\alpha_2, \alpha_1+2\alpha_2, -\alpha_1-\alpha_2,$ and $-\alpha_1-2\alpha_2$.  In all cases, it is the $i=2$ condition that is non-trivial, and the roots are second-lowest or second-highest in $\alpha_2$-strings of 4 roots:
\ben\label{fourstrings} 
\{\alpha_1+3\alpha_2, \alpha_1+2\alpha_2, \alpha_1+\alpha_2, \alpha_1 \}\, ,\ \ \{ -\alpha_1, -\alpha_1-\alpha_2, -\alpha_1-2\alpha_2, -\alpha_1-3\alpha_2 \}\ .
\een
This makes sense if we recall the Weyl-group algorithm for fusion described in \cite{Kac90, MW90}, along with its relation to the depth rule \cite{MW94}.  The refined depth rule (\ref{adepthfsymm}) tells us, first,  that a state (or vector) of the appropriate weight must exist. However, a `cancelling state' (or vector) may also be present.  The first state and the cancelling state have weights related by the shifted action of a Weyl group element, the relevant primitive reflection $r_i$ ($i\in \hat I$).  In our examples, it is the state with second-lowest or second-highest weight in the $\alpha_2$-string ($\alpha_1+\alpha_2, -\alpha_1 -2\alpha_2$; or $\alpha_1+2\alpha_2, -\alpha_1-\alpha_2$) that can be eliminated by a cancelling state of lowest weight ($\alpha_1$ or $-\alpha_1-3\alpha_2$), respectively.

\vskip.5cm
\begin{table}[!ht]
\begin{center}
    \begin{tabular}{ | l | l | l | l | l | l | l |}
    \hline
    root $\nu-\mu$ & $\mu_0\geq\quad$ & $\mu_1\geq\quad$ & $\mu_2\geq\quad$  & $\nu_0$ & $\nu_1$ & $\nu_2$ \\ \hline\hline
   $\alpha_1$ &\  1 &\  0 &\  3& $\mu_0-1\quad$ & $\mu_1+2\quad$ & $\mu_2-3\quad$\\ \hline
   $\alpha_1+\alpha_2$ &\  1 &\  0 &\  2\ \ \ *& $\mu_0-1\quad$ & $\mu_1+1\quad$ & $\mu_2-1\quad$ \\ \hline
    $\theta=2\alpha_1+3\alpha_2$ &\  2 &\  0 &\  0& $\mu_0-2\quad$ & $\mu_1+1\quad$ & $\mu_2\quad$ \\ \hline 
   $\alpha_1+2\alpha_2$ &\  1 &\  0&\  1\ \ \ * & $\mu_0-1\quad$ & $\mu_1\quad$ & $\mu_2+1\quad$ \\ \hline
    $\alpha_1+3\alpha_2$ &\  1 &\  1 &\  0& $\mu_0-1\quad$ & $\mu_1-1\quad$ & $\mu_2+3\quad$ \\ \hline
    $\alpha_2$ &\  0 &\  1 &\  0& $\mu_0\quad$ & $\mu_1-1\quad$ & $\mu_2+2\quad$ \\ \hline 
     $-\alpha_1$ &\  0 &\  2 &\  0& $\mu_0+1\quad$ & $\mu_1-2\quad$ & $\mu_2+3\quad$ \\ \hline
    $-\alpha_1-\alpha_2$ &\  0 &\  1 &\  1\ \ \ *& $\mu_0+1\quad$ & $\mu_1-1\quad$ & $\mu_2+1\quad$ \\ \hline
    $-\theta=-2\alpha_1-3\alpha_2$ &\  0 &\  1 &\  0& $\mu_0+2\quad$ & $\mu_1-1\quad$ & $\mu_2\quad$ \\ \hline 
   $-\alpha_1-2\alpha_2$ &\  0 &\  0 &\  2\ \ \ *& $\mu_0+1\quad$ & $\mu_1\quad$ & $\mu_2-1\quad$ \\ \hline
    $-\alpha_1-3\alpha_2$ &\  0 &\  0 &\  3& $\mu_0+1\quad$ & $\mu_1+1\quad$ & $\mu_2-3\quad$ \\ \hline
     $-\alpha_2$ &\  0 &\  0 &\  2& $\mu_0\quad$ & $\mu_1+1\quad$ & $\mu_2-2\quad$ \\ \hline  
    \end{tabular}
   \vskip.5cm\caption{\label{G2table}$G_2$ off-diagonal adjoint fusion. Conditions from (\ref{OffDiagF}) for nonzero $G_2$ off-diagonal adjoint fusion coefficient ${}^{(k)}N_{\theta,\mu}^\nu=1$ are displayed in columns 2, 3 and 4. The last 3 columns give the affine Dynkin labels for the weight $\hat\nu$.  The only non-trivial constraints are indicated by *.  All others are automatically satisfied if $\mu, \nu\in P_+^k$. }
\end{center}
\end{table}

Among representations, the adjoint is not large.  Consequently, adjoint fusion can only be a little off-diagonal.  Furthermore, few cancellations can occur; few depth-rule constraints are non-trivial in the sense just described.  

Most importantly, consider the `affine' (or non-finite) constraints of the refined depth rule  (\ref{adepthfsymm}, \ref{adepthesymm}), meaning the $i=0$ ones. 
Table~\ref{G2table} shows that for $G_2$, they are trivial -- as long as the weights $\mu, \nu\in P_+^k$, they are satisfied. 

We will now show that this is a general feature of non-diagonal adjoint fusion.  Let us treat the 2 cases $\nu-\mu\in \Delta_\pm$ separately.  First, we write a formula for the off-diagonal tensor product multiplicities $c_{\theta, \mu}^{\nu\not=\mu}$ by rewriting (\ref{OffDiagF}) and dropping the extra, `affine' constraint:  
\ben    c_{\theta, \mu}^{\nu\not=\mu}\ =\ 
  \begin{cases}
      1\, , & \text{if}\ \nu-\mu\in \Delta_+, \ \nu-\mu+(\mu_i+1)\alpha_i  \, \not\in \Delta_+\, \ \forall\, i\in I\ ;\\ 
     1\, , & \text{if}\ \nu-\mu\in \Delta_-,  \ \nu-\mu-(\nu_i+1)\alpha_i \, \not\in \Delta_-\, \ \forall\, i\in I\ ;\\ 
      0\, ,  &  \text{otherwise}\, . 
  \end{cases}  
\label{OffDiagc}
\een
Now consider the further conditions that would be imposed to find the fusion coefficients ${}^{(k)}N_{\theta, \mu}^{\nu\not=\mu}$: 
\begin{align}\label{OffDiagFNzero}
     \nu-\mu\in \Delta_+:& \ \ \ \ \nu-\mu +(\nu_0+1)\theta\ \not\in \Delta_+\ ,\\ 
     \nu-\mu\in \Delta_-:&\ \ \ \ \nu-\mu -(\mu_0+1)\theta\ \not\in \Delta_-\ .
  \end{align}
  Since $\theta$ is the highest root, however, these conditions are always satisfied, if $\mu, \nu \in P_+^k$.  Dropping them, we see that 
\ben  
{}^{(k)}N_{\theta, \mu}^{\nu\not=\mu}\ =\ c_{\theta, \mu}^{\nu\not=\mu}\ , \ \  \text{for}\  \theta,\mu,\nu\in P_+^k\ .
\label{NcOff}\een 

Since $\alpha_0=-\theta$, the triviality of the affine $i=0$ depth-rule condition for off-diagonal adjoint fusion may be understood in terms of $\theta$-strings of weights in the adjoint representation. Since $\theta$ is the unique highest root, the unique longest $\theta$-string is $\{\theta, 0, -\theta\}$.  It only causes cancellations in the diagonal adjoint fusion case, however.  $\theta$-strings with 1 or 2 elements cannot give rise to the same effect. 

Eqn.~(\ref{OffDiagc}) coincides with the `Adjoint Rule' found by Stembridge in \cite{Stem}, off-diagonal part.  To see this, first rewrite (\ref{OffDiagc}) as
\ben    c_{\theta, \mu}^{\nu\not=\mu}\ =\ 
  \begin{cases}
      1\, , & \text{if}\ \nu-\mu\in \Delta,\  \ \nu-\mu+(\mu_i+1)\alpha_i  \, \not\in \Delta\cup\{0\}\, ,\ \ \forall\, i\in I\ ;\\  
      0\, ,  &  \text{otherwise}\, . 
  \end{cases}  
\label{OffDiagcTwoCase}
\een   
This is just eqn.~(\ref{OffDiagF}), minus the affine $i=0$ condition.  For each root $\beta\in \Delta$, an $\alpha_i$-depth $d_i[\beta]$ can be defined; it is the maximum integer $u$ such that $\beta+u\alpha_i\in \Delta$.  Then the $i$-th condition of (\ref{OffDiagcTwoCase}) is just $\mu_i\geq d_i[\nu-\mu]$. Following \cite{Stem}, apart from a change in sign, we define the depth (weight) 
\ben\label{DepthWeight}
\delta[\beta]\ :=\ \sum_{j\in I}\,  d_j[\beta]\, \Lambda^j\ ,
\een
for $\beta\in\Delta$.  Then the $r$ conditions $\mu_i\geq d_i[\nu-\mu]$, $i\in I$,  can be combined into $\mu - \delta[\nu-\mu]\in P_+$. We reproduce  the off-diagonal part of the Adjoint Rule, Prop.~2.14 in \cite{Stem}: 
\ben    c_{\theta, \mu}^{\nu\not=\mu}\ =\ 
  \begin{cases}
      1\, , & \text{if}\ \nu-\mu\in \Delta,\  \ \mu-\delta[\nu-\mu]\in P_+\ \, ;\\  
      0\, ,  &  \text{otherwise}\, . 
  \end{cases}  
\label{OffDiagcStem}
\een  

For computations, we prefer eqn.~(\ref{OffDiagc}) over (\ref{OffDiagcStem}). The separation into cases $\nu-\mu\in\Delta_\pm = \pm\Delta_+$ is helpful, and many of the $r$ conditions are trivial for certain $i\in I$. 

By (\ref{OffDiagcTwoCase}), for nontriviality we require that 
\ben\label{NonTrivI} 
\nu-\mu \ =:\  \beta \in \Delta,\  \ \nu-\mu+(\mu_i+1)\alpha_i  \, \in \Delta\cup\{0\}\, ,\ \ {\text{for\ some\ }} i\in I\  \een
be nontrivial and stronger than the requirement $\mu, \nu\in P_+$. Consider the $\alpha_i$-string of roots containing $\beta=\nu-\mu$.  
Recall that  $d_i[\beta]$ is the $\alpha_i$-depth of a root $\beta\in \Delta$: it is the maximum positive integer $u$ such that   $\beta+u\alpha_i\in \Delta_+$.  Eqn.~(\ref{NonTrivI} ) is just $\mu_i\leq d_i[\beta]$. On the other hand, requiring $\nu=\mu+\beta\in P_+$ demands $\mu_i\geq -\beta_i= -(\beta,\alpha^\vee_i)$. Nontriviality is possible 2 ways: 
\ben\label{NonTrivII}
1)\ \beta_i\leq 0\ \ \&\ \ d_i[\beta]>-\beta_i\ ,\ \ \ {\text{ or}}\ \ \ \ 2) \ \beta_i>0\ \ \&\ \ d_i[\beta]>0\ . 
\een 
For example, the 4 boxes marked * in Table~\ref{G2table} refer to cases of type 1,2,2, and 1, in that order.  Notice that both cases of (\ref{NonTrivII}) require a positive depth $d_i[\beta]$; they can be combined as 
\ben\label{NonTrivComb}
d_i[\beta]\ >\ \max(0, -\beta_i)\ \ . 
\een

Define the $\alpha_i$-height $h_i[\beta]$ as the maximum integer $v$ such that $\beta-v\alpha_i\in \Delta$.  Since $h_i[\beta]-d_i[\beta] = \beta_i$, we can rewrite (\ref{NonTrivII}) more symmetrically as 
\ben\label{NonTrivIII}
1)\ \beta_i\leq 0\ \ \&\ \ h_i[\beta]>0\ ,\ \ \ {\text{ or}}\ \ \ \ 2) \ \beta_i>0\ \ \&\ \ d_i[\beta]>0\ . 
\een 
The root $\beta$ cannot be the highest or lowest in the $\alpha_i$-string. The string should therefore be ``long'' in the terminology of \cite{Stem}, having 3 or more elements.  For off-diagonal tensor products, the `degenerate' strings $\{ \alpha_i, 0, -\alpha_i \}$ are irrelevant.   According to \cite{Stem}, no string has more than 4 elements, and long non-degenerate strings occur only if the algebra has both long
and short roots. Furthermore, the highest and lowest roots in such a string must be long, and the interior roots and $\alpha_i$ must be short. The nontrivial $G_2$ examples of Table~\ref{G2table} are precisely of this type.  Another useful observation made in \cite{Stem} is that for every $\beta\in \Delta$, there is at most one
index $i$ such that $d_i[\beta]>0$ and the $\alpha_i$-string through $\beta$ is long.  A  table for any algebra $X_r$ like Table~\ref{G2table} for $G_2$ would also have a maximum of 1 * in each row. 

These considerations show that most of the conditions of (\ref{OffDiagc}) are trivial, making computations less complicated than might be expected. For example, as we show in  Appendix A, all $r$ conditions are trivial when the algebra is simply-laced ($X= A, D, E)$, so that adjoint fusion is very simple.    Off-diagonal adjoint tensor product (fusion) is like the Pieri rule for $A_r$ fundamental representations: there are no cancellations --  one only drops representations with weights not in $P_+$ ($P_+^k$).  In the same Appendix, we also analyze the remaining algebras and classify the positive roots giving rise to nontrivial conditions for off-diagonal adjoint fusion/tensor products. They are listed there in Table~\ref{nontrivialConditions}.

To summarize, only the diagonal adjoint fusion coefficients are reduced from the corresponding tensor product coefficients, according to the simple rule (\ref{DiagAdjFusDLs}).  The off-diagonal coefficients are identical.  The affine depth rule condition, the $i=0$ one,  is trivial except for the diagonal coefficients. 

The off-diagonal coefficients can be computed using the universal formulas  (\ref{OffDiagc}) or (\ref{OffDiagcTwoCase}), both equivalent to (\ref{OffDiagcStem}), previously derived in \cite{Stem}. At the expense of universality, however, one can calculate much more efficiently, using Table~\ref{nontrivialConditions} of Appendix A.

\section{Adjoint Fusion Tadpoles}

The fusion tadpole ${}^{(k)}T_\lambda$ is the 1-point genus-1 fusion dimension. It can be written as the sum over diagonal fusion coefficients, or as the trace of the corresponding fusion matrix.  For example,  the adjoint fusion tadpole is 
\ben {}^{(k)}T_\theta \ =\ \sum_{\mu\in P_+^k}\, {}^{(k)}N_{\theta,\mu}^\mu\ .  \label{adjTsum}\een 
The simple formula (\ref{DiagAdjFusDLs}) for diagonal adjoint fusion means that the adjoint affine tadpole can be calculated as a sum over the lattice polytope $P_+^k$ of a simple integer function, the number of nonzero Dynkin labels of the corresponding affine weight, minus 1.  Geometrically, these numbers equal the dimensions of the (lattice-polytope) faces of the polytope $P_+^k$ to which the points belong. 

For each of the Lie algebras it is possible to write down a set of polynomial formulas for the adjoint fusion tadpoles. We compute them for the  classical algebras  $A_r, B_r, C_r, D_r$, as well as for $E_6$. The last case shows that our methods work for the exceptional algebras, but also indicates the large number of equations they require.

To compute ${}^{(k)}T_\theta$, we reduce the problem to smaller combinatorial problems. We will also need $\dim P_+^k$, equal to the zero fusion tadpole ${}^{(k)}T_0$ for the identity primary field, related to the scalar representation of $X_r$: 
\ben  {}^{(k)}T_0\ =\ \dim P_+^k\ =\ \dim\Big\{ \lambda\in P \,\Big\vert\, \sum_{j\in \hat I}m^\vee_j\lambda_j = k,\, \lambda_j\in\N_0 \Big\} \ .\label{zeroT}\een 
These values are well known, but we will compute a couple of them here as simple illustrations of the techniques that helped us compute the tadpoles.  For the adjoint tadpole, we sometimes find it  simpler to avoid the -1 in (\ref{DiagAdjFusDLs}), and deal with 
\ben\label{TthzDef}
{}^{(k)}T_{\theta+0}\ :=\ {}^{(k)}T_{\theta} + {}^{(k)}T_{0}\ =\ \sum_{\mu\in P_+^k}\, \dim\left\{\, \mu_j\not=0,\  j\in\hat I  \,\right\}\ .
\een

The main difficulty in the computations arises from differing values of the co-marks $m^\vee_i, i\in\hat I$.  The number of formulas required (or pieces for the piecewise-polynomial result) will be ${\mathrm{lcm}}(m_1^\vee, \ldots, m_r^\vee)$ \cite{HP09}.  

\subsection{Sample Tadpole Calculation: $\mathbf{A_{r,k}}$ and $\mathbf{C_{r\geq2,k}}$}

One way to reduce the computation of ${}^{(k)}T_\theta$ is to partition  the set of Dynkin labels into smaller sets with the same value co-mark, and then apply the formulas for $A_{r,k}$, for which $m^\vee_j=1\ \forall i\in\hat I$.\footnote{\, Essentially, this is the strategy used in \cite{HP09} to calculate $\dim P_+^k$ for different Lie algebras.}  In order to proceed, then, we need to work out our results for $A_{r,k}$ first.  

Incidentally, the same formulas work for both $A_{r,k}$ and $C_{r,k}$, since they invoke identical  combinatorial problems: the co-marks of both algebras are all 1.  

We will make use of the falling power notation 
\begin{align}
(\ell +c)^{\underline{k}}\ :=\  (\ell +c)(\ell +c-1) \dots (\ell +c-k+1) \ ,
\end{align}
with $(\ell +c)^{\underline{0}}=1$, because of the identities
\begin{align}
\label{eq:sumIdentity}
\sum_{\ell=L_1}^{L_2} (\ell +c)^{\underline{m}}\ =\ \frac{(L_2+1 + c)^{\underline{m+1}}}{m+1}-\frac{(L_1+c)^{\underline{m+1}}}{m+1}\ 
\end{align}
and 
\ben 
\Delta_\ell\, (\ell+c)^{\underline{m}}\ =\ m (\ell+c)^{\underline{m-1}}\ .
\label{DeltaFP}\een 
Here $\Delta_\ell$ is the forward difference operator:
\ben
\Delta_\ell f(\ell) \ :=\ f(\ell+1)-f(\ell)\ .
\label{Deltal}\een 

To calculate the zero tadpole ${}^{(k)}T_0$ for $A_r$, or 
\ben\label{TzA}
T_0[A_{r,k}]\ =\ \dim\Bigg\{ \sum_{j\in\hat I} \lambda_j \hat\Lambda^j \ \Big\vert\ \sum_{j\in\hat I} \lambda_j = k\, ,\ \lambda_j\in \N_0 \Bigg\}\ ,
\een 
we fix one affine Dynkin label ($\lambda_0$, say, to $u$) to relate it to  
$T_0[A_{r-1,k-u}]$.  With $s=k-u$, we obtain 
\ben\label{TzAsum}
T_0[A_{r,k}]\ =\ \sum_{s=0}^k T_0[A_{r-1,s}]\ , 
\een
or, equivalently, 
\ben\label{TzAdiff}
\Delta_k T_0[A_{r,k}]\ =\ T_0[A_{r-1,k+1}]\ . 
\een
The formula 
\ben  T_0[A_{r,k}]\ =\ \frac{(k+r)^{\underline{r}}}{r!}\ \label{zeroDimA}. \een
can be proved easily by induction, and verified by showing that (\ref{TzAsum}, \ref{TzAdiff}) are satisfied, using (\ref{eq:sumIdentity}, \ref{DeltaFP}), respectively. 

A similar computation works for the adjoint tadpole.  Again imagine fixing one of the Dynkin labels, $\lambda_0$, say.  Now, because of eqn.~(\ref{DiagAdjFusDLs}), the cases $\lambda_0=0$ and $\lambda_0\not=0$ must be treated differently.  We find 
\ben\label{TthSumAz}
T_\theta[A_{r,k}]\ =\ T_\theta[A_{r-1,k}]\ +\ \sum_{s=0}^{k-1} T_{\theta+0}[A_{r-1,s}]\ ,
\een
or, using (\ref{zeroDimA}), 
\ben\label{TthSumA}
T_\theta[A_{r,k}]\ =\  \sum_{s=0}^{k} T_\theta[A_{r-1,s}]\  +\ \frac{(k+r-1)^{\underline{r}}}{r!}\ .
\een
Equivalently, we can write
\ben\label{DelTthA}
\Delta_k\, T_\theta[A_{r,k}]\ =\ T_\theta[A_{r-1,k+1}]\ +\ \frac{(k+r-1)^{\underline{r-1}}}{(r-1)!}\ .
\een
Induction with eqn.~(\ref{TthSumA}) leads to 
\ben\label{TthA}
T_\theta[A_{r,k}]\ =\ (k-1)\,\frac{(k+r-1)^{\underline{r-1}}}{(r-1)!}\ .
\een 
Using eqns.~(\ref{eq:sumIdentity})  and (\ref{DeltaFP}), we can verify eqns.~(\ref{TthSumA}) and (\ref{DelTthA}), respectively.

\subsection{Sample Tadpole Calculation: $\mathbf{B_{r,k} = B_{r,2J+1}}$,  $\mathbf J\in\N$}

We consider the Lie algebras of type $B_r$ as an example of how the formulas for adjoint fusion tadpoles are determined, when the co-marks are not all 1. As a first step, let us calculate ${}^{(k)}T_0\,=\, T_0[B_{r,k}]$. 

In $B_r$, the `tip' co-marks $m^\vee_0=m^\vee_1=m_r^\vee = 1$, while the rest, the `interior' ones,  have value $m^\vee_2=m^\vee_3=\cdots =m^\vee_{r-1}= 2$.  We need to count the number of solutions to $\sum_{j\in \hat I} m_j^\vee \lambda_j = k$, with $\lambda_j\in \N_0$ for all $j\in \hat I$. 

Suppose that one of the interior Dynkin labels is fixed. Since it doesn't matter which, set  $\lambda_2=u \leq k/2$. Then that part of the problem is reduced to calculating $T_0[B_{r-1,k-2u}]$. Because the relevant co-mark $m_2^\vee=2$, it becomes clear that we need to treat odd and even levels $k$ separately. Let us consider the odd case: $k=2J+1$,  $J\in\N$.

By considering all possible values of the Dynkin label $\lambda_1$, we can write 
\ben T_0[B_{r,2J+1}]\ =\ \sum_{s=0}^J T_0[B_{r-1,2s+1}]\ .\label{TzeroSum}\een
Alternatively, with $\Delta_J$ the forward difference operator, we have \ben \Delta_J\, T_0[B_{r,2J+1}]\ =\ T_0[B_{r,2(J+1)+1}]  - T_0[B_{r,2J+1}] \ =\ T_0[B_{r-1,2(J+1)+1}]\ .  
\label{DeltaJTz}\een

By calculating cases with small rank and level, it is not difficult to find the expression 
\ben\label{TzeroB}
T_0[B_{r,2J+1}]\ =\ 3 \frac{(J+r)^{\underline{r}}}{r!}+ \frac{(J+r-1)^{\underline{r}}}{r!}
\een
This formula can be proved  by induction using (\ref{TzeroSum}). 
Alternatively,  its validity can be confirmed by using  (\ref{DeltaFP}) to show that (\ref{DeltaJTz}) is satisfied. 

A similar method works for the tadpole $T_\theta[B_{r,2J+1}]$. Eqn. (\ref{DiagAdjFusDLs}) tells us that we need to not only count the elements of $P_+^k$, but also weight them with their numbers of nonzero Dynkin labels. 

Consider the contributions to  $T_\theta[B_{r,2J+1}]$ with one co-mark-2 Dynkin label fixed, $\lambda_2$ say.  Now we must treat the cases $\lambda_2=0$ and $\lambda_2>0$ differently. When $\lambda_2=0$, the problem reduces to the same problem, but for rank $r-1$: the contribution is just $T_\theta[B_{r-1,2J+1}]$.  However, when $\lambda_2=u>0$, the level is reduced as well as the rank, and the nonzero $\lambda_1$ produces an addition of 1 for each weight of $B_{r-1, 2(J-u)+1}$: so the contribution is $T_\theta[B_{r-1,2(J-u)+1}] + T_0[B_{r-1,2(J-u)+1}]$. 

The analogs of (\ref{TzeroSum}) and (\ref{DeltaJTz}) for the tadpole are therefore 
\ben    
T_\theta[B_{r,2J+1}]\ =\ \sum_{s=0}^J T_\theta[B_{r-1, 2s+1}]\ +\ \sum_{s=0}^{J-1} T_0[B_{r-1, 2s+1}]\ 
\label{TthetaSum}\een
and 
\ben
\Delta_J\, T_\theta[B_{r,2J+1}]\ =\ T_\theta[B_{r-1, 2(J+1)+1}]\ -\ T_0[B_{r-1, 2J+1}]\ , 
\label{DeltaJTtheta}\een 
respectively.  As for $T_0[B_{r,k}]$, calculations at small rank and level lead to a formula
\ben  
T_\theta[B_{r,2J+1}]\ =\ 4\frac{(J+r-1)^{\underline{r}}}{(r-1)!}\ -\ (r-2) \frac{(J+r-2)^{\underline{r-1}}}{(r-1)!}\ .
\label{TthetaB}\een 
This result can be confirmed in either of (\ref{TthetaSum}) or (\ref{DeltaJTtheta}).  Table~\ref{adjointValuesB} displays $T_\theta[B_{r,2J+1}]$ values for $r, J\leq  6$.

\vskip.5cm
\begin{table}[!ht]
\begin{center}
    \begin{tabular}{ | l || l | l | l | l |}
    \hline
 \diagbox{$k$}{$r$} &  3 & 4  &  5 & 6\\ \hline\hline
   2 &\  3 &\  3 &\  3 &\  3  \\ \hline
   3 &\  11 &\  14 &\  17 &\  20  \\ \hline
   4 &\  24 &\  34 &\  45 &\  57  \\ \hline 
   5 &\  45 &\  72 &\  105 &\  144  \\ \hline
   6 &\  74 &\  130 &\  205 &\  301  \\ \hline
   7 &\  114 &\  220 &\  375 &\  588  \\ \hline 
   8 &\  165 &\  345 &\  630 &\  1050  \\ \hline
   9 &\  230 &\  520 &\  1015 &\  1792  \\ \hline
  10 &\  309 &\  749 &\  1554 &\  2898  \\ \hline 
  11 &\  405 &\  1050 &\  2310 &\  4536  \\ \hline
  12 &\  518 &\  1428 &\  3318 &\  6846  \\ \hline
  13 &\  651 &\  1904 &\  4662 &\  10080  \\ \hline  
    \end{tabular}
\vskip.5cm\caption{\label{adjointValuesB} Values of the adjoint fusion tadpoles $T_\theta [B_{r,k}]$.}
\end{center}
\end{table}


Another strategy is to treat the $B_{r,k}$ Dynkin labels as  two sets of $A_{r,k}$ labels, one set of 3 labels with co-mark 1, the other with $r-2$ co-mark-2 labels.  Then the $A_{r,k}$ results (\ref{zeroDimA}, \ref{TthA}) can be used. 

The zero tadpole satisfies
\ben\label{TzBrJ} T_0[B_{r,2J+1}]\ =\ \sum_{s=0}^J T_0[A_{2,2(J-s)+1}]\, T_0[A_{r-3,s}]\ . \een
This formula can be used to find $T_0[B_{r, 2J+1}]$ by induction. 

To treat the adjoint tadpole, we recall eqn.~(\ref{TthzDef}). We deduce 
\ben\label{tTBrJ} T_{\theta+0}[B_{r,2J+1}]\ =\ \sum_{s=0}^J \Big\{\, T_{\theta+0}[A_{2,2(J-s)+1}]\, T_0[A_{r-3,s}]  \ +\  
T_0[A_{2,2(J-s)+1}]\, T_{\theta+0}[A_{r-3,s}]  \,\Big\}\ . \een  
The result (\ref{TthetaB}) can also be found from this description.

\subsection{Adjoint fusion tadpoles for all classical algebras, and $E_6$ }

Via  procedures similar to those just illustrated for $B_{r,2J+1}$, we determined the following formulas for the adjoint fusion tadpoles of the classical algebras, plus $E_6$.

\noindent $\mathbf {A_{r},\  r \geq 1:}$
\begin{align}
T_\theta[A_{r,k}] =&  \frac{(k + r-1)^{\underline{(r-1)}}}{(r-1)!}(k-1).
\end{align}
$\mathbf{B_{r},\ r \geq 3:}$
\begin{align}
\label{eq:tadpoleB}
T_\theta[B_{r,k}] =& \begin{cases}
4\frac{(J+r-1)^{\underline{r}}}{(r-1)!}-3(r-1)\frac{(J+r-2)^{\underline{r-1}}}{(r-1)!} -\frac{(J+r-1)^{\underline{r-1}}}{(r-1)!}, & {\mathrm  {if}\ } k =2J,\\
4\frac{(J+r-1)^{\underline{r}}}{(r-1)!}-(r-2)\frac{(J+r-2)^{\underline{r-1}}}{(r-1)!}, & {\mathrm{if}\ } k = 2J+1.
\end{cases}
\end{align}
\noindent$\mathbf{C_{r}, r \geq 2:}$
\begin{align}
T_\theta[C_{r,k}] =&  \frac{(k + r-1)^{\underline{(r-1)}}}{(r-1)!}(k-1).
\end{align}
$\mathbf{D_{r}, r \geq 4:}$
\begin{align}
T_\theta[D_{r,k}] =& 
\begin{cases}
8J\frac{(J+r-2)^{\underline{r-1}}}{(r-1)!}+(r-4)\frac{(J+r-3)^{\underline{r-2}}}{(r-2)!}-\frac{(J+r-3)^{\underline{r-3}}}{(r-3)!},& {\mathrm{if}\ } k = 2J,\\
8\frac{(J+r-2)^{\underline{r}}}{(r-1)!} +4r\frac{(J+r-2)^{\underline{r-1}}}{(r-1)!}, &{\mathrm{if}\ } k =2J+1.\\
\end{cases}
\end{align}
$\mathbf{E_6:}$
\begin{align}
\label{eq:tadpoleE6}
T_\theta[E_6] =& 
\begin{cases}
\frac{81}{5}J^6 + \frac{324}{5}J^5 + \frac{189}{2}J^4 +\frac{117}{2}J^3 +\frac{54}{5}J^2 - \frac{14}{5} J -1, &{\mathrm{if}\ } k = 6J,\\
\frac{81}{5}J^6 + 81 J^5 + \frac{621}{4}J^4 +141 J^3 +\frac{1191}{20}J^2 +9 J, &{\mathrm{if}\ } k = 6J + 1,\\
\frac{81}{5}J^6 + \frac{486}{5}J^5 + \frac{459}{2}J^4 +\frac{537}{2}J^3 +\frac{1593}{10}J^2 + \frac{423}{10} J +3, &{\mathrm{if}\ } k = 6J+2,\\
\frac{81}{5}J^6 + \frac{567}{5}J^5 + \frac{1269}{4}J^4 +450 J^3 +\frac{6741}{20}J^2 + \frac{1241}{10} J +17, &{\mathrm{if}\ } k = 6J+3,\\
\frac{81}{5}J^6 + \frac{648}{5}J^5 + \frac{837}{2}J^4 +\frac{1389}{2}J^3 +\frac{3099}{5}J^2 + \frac{1392}{5} J +48, &{\mathrm{ if}\ } k = 6J+4,\\
\frac{81}{5}J^6 + \frac{729}{5}J^5 + \frac{2133}{4}J^4 +1011 J^3 +\frac{20871}{20}J^2 + \frac{2766}{5} J +117, &{\mathrm{if}\ } k = 6J+5.\\
\end{cases}
\end{align}
Here $J\in\N_0$ is assumed. These formulas were checked to agree with tadpole values computed from the Verlinde formula \cite{Verlinde88} using LieART \cite{FK15} up to rank $r=7$ and level $k=18$. 

One can obtain piecewise polynomial formulas for the adjoint fusion tadpoles of the remaining exceptional algebras, most simply by computer.  However, as noted above, the number of pieces (polynomial formulas) required  would be ${\mathrm {lcm}}(m_1^\vee, \dots, m_{N-1}^\vee)$. For $E_7$ and $E_8$, e.g.,  the numbers are  $12$ and $60$, respectively.  We refrain from completing the list started above. 

\subsection{Zero fusion tadpoles for all classical algebras, and $E_6$.} The zero tadpoles were also found and are listed below, for the reader's convenience.

\noindent $\mathbf {A_{r},\  r \geq 1:}$
\begin{align}
T_0[A_{r,k}] = \frac{(k + r)^{\underline{r}}}{r!},
\end{align}
\noindent $\mathbf {B_{r},\  r \geq 3:}$
\begin{align}
T_0[B_{r,k}]\ =\ \begin{cases} 
\frac{(J+r)^{\underline{r}}}{r!}+3 \frac{(J+r-1)^{\underline{r}}}{r!}\ & \text{\rm if } k = 2J,   \\
3 \frac{(J+r)^{\underline{r}}}{r!}+ \frac{(J+r-1)^{\underline{r}}}{r!}\  & \text{\rm if } k = 2J+1, 
\end{cases}
\end{align}
\noindent $\mathbf {C_{r},\  r \geq 2:}$
\begin{align}
T_0[C_{r,k}]= \frac{(k + r)^{\underline{r}}}{r!},
\end{align}
\noindent $\mathbf {D_{r},\  r \geq 4:}$
\begin{align}
T_0[D_{r,k}] = \begin{cases}
8\frac{(J+r-1)^{\underline{r}}}{r!}+\frac{(J+r-2)^{\underline{r-2}}}{(r-2)!}\ , & \text{\rm if } k = 2J,\\
8\frac{(J+r-1)^{\underline{r}}}{r!}+4\frac{(J+r-1)^{\underline{r-1}}}{(r-1)!}\ , & \text{\rm if } k =2J+1.\\
\end{cases}
\end{align}
$\mathbf{E_6:}$
\begin{align}
\label{eq:alcoveE6}
T_0[E_6] =& 
\begin{cases}
\frac{27}{10}J^6 + \frac{81}{5}J^5 + \frac{153}{4}J^4 +45 J^3 +\frac{551}{20}J^2 + \frac{83}{10} J +1, &{\mathrm{if}\ } k = 6J,\\
\frac{27}{10}J^6 + \frac{189}{10} J^5 + \frac{423}{8}J^4 +\frac{301}{4} J^3 +\frac{2277}{40}J^2 +\frac{427}{20} J+3, &{\mathrm{if}\ } k = 6J + 1,\\
\frac{27}{10}J^6 + \frac{108}{5}J^5 + \frac{279}{4}J^4 +116 J^3 +\frac{2091}{20}J^2 + \frac{242}{5} J +9, &{\mathrm{if}\ } k = 6J+2,\\
\frac{27}{10}J^6 + \frac{243}{10}J^5 + \frac{711}{8}J^4 +\frac{675}{4} J^3 +\frac{6997}{40}J^2 + \frac{1869}{20} J +20, &{\mathrm{if}\ } k = 6J+3,\\
\frac{27}{10}J^6 + 27 J^5 + \frac{441}{4}J^4 +235 J^3 +\frac{5511}{20}J^2 + \frac{337}{2} J +42, &{\mathrm{ if}\ } k = 6J+4,\\
\frac{27}{10}J^6 + \frac{297}{10}J^5 + \frac{1071}{8}J^4 +\frac{1265}{4} J^3 +\frac{16497}{40}J^2 + \frac{5621}{20} J +78, &{\mathrm{if}\ } k = 6J+5.\\
\end{cases}
\end{align}

\section{Conclusion}

Let us conclude by pointing out the main results of this note. A simple and universal description of adjoint affine fusion is given by  (\ref{DiagAdjFusDLs}, \ref{NcOff}).  Equation (\ref{DiagAdjFusDLs}) determines the fusion coefficient ${}^{(k)}N_{\theta,\mu}^\mu$ for diagonal adjoint fusion as the number of nonzero (affine) Dynkin labels of $\hat\mu$ minus 1.  This gives it a nice polytope interpretation. Decompose the lattice polytope defined by $P_+^k$ into its faces.  Then ${}^{(k)}N_{\theta,\mu}^\mu$ is  the dimension of the face containing $\mu\in P_+^k$.  

The universal description of adjoint affine fusion is completed by (\ref{OffDiagF}).  The off-diagonal coefficient ${}^{(k)}N_{\theta,\mu}^{\nu\not=\mu}$ is either 1 or 0.  It is 1 iff 1) $\nu-\mu$ is a root, and 2) $\nu-\mu+(\mu_i+1)\alpha_i$ is not zero nor a root, for all $i\in \hat I$.  As a consequence, the off-diagonal adjoint fusion coefficient equals the corresponding tensor product coefficient, as stated in eqn.~ (\ref{NcOff}). Hence eqn.~(\ref{OffDiagc}) computes both efficiently. 

But this efficiency can be improved greatly -- most of the conditions of (\ref{OffDiagc}) can be safely ignored.  Only short, simple roots need to be considered, and for those, only a small number of possibilities $\nu-\mu=\beta\in \Delta$ must be checked. We found those possibilities, and list them in Table~\ref{nontrivialConditions} of  Appendix A.

The result is a pleasingly minimal adaptation to affine fusion of the Adjoint Rule of \cite{Stem} for tensor products.  Diagonal adjoint fusion is simple, and the off-diagonal part is remarkably Pieri-like. 

Due to the importance of the adjoint representation, we believe these results will be useful.\footnote{\,We note that some previous work on adjoint tensor products was motivated by their use in the cohomology of Lie algebras \cite{Kostant97, BZ89}.   Although  affine fusion is relevant to the quantum cohomology of flag manifolds \cite{WittenQCoh, Fulton04, MS12}, we are unaware of a role that adjoint fusion might play in some `quantum cohomology' of Lie algebras.} As one application, the formula (\ref{DiagAdjFusDLs}) allowed us to calculate the genus-1 1-point adjoint fusion dimension, or adjoint fusion tadpole, for all classical Lie algebras, and $E_6$.  The expressions obtained are piecewise-polynomial in the level, and are listed in subsection 4.3. Arguably, fusion tadpoles are interesting as the next-to-simplest type of nonzero-genus fusion dimension. 

\appendix\section{Nontrivial off-diagonal adjoint conditions}

Table~\ref{G2table} treats $G_2$, and we are interested in classifying the analogs for all simple Lie algebras of its boxes marked by *.   Here we treat the remaining algebras using the descriptions of their root systems found in \cite{Bour}. That is, for $X_r = A_r, B_r, C_r, D_r, E_6, E_7, E_8, F_4$, we find all the roots $\beta\in\Delta(X_r)$ and their indices $i\in I$, such that if $\nu-\mu=\beta$, the condition of (\ref{OffDiagcTwoCase}) is nontrivial, i.e.~not automatically satisfied when  $\mu, \nu\in P_+$.  By \cite{Stem}, we know that such a $\beta$ must be part of a long, nondegenerate string of roots, and that there is at most 1 such index $i$ for each $\beta$. 

Necessary conditions for nontriviality were given above: (\ref{NonTrivII}, \ref{NonTrivComb}) and (\ref{NonTrivIII}) are equivalent. 

Half of the 4 nontrivial $G_2$ conditions of Table~\ref{G2table} involve $\nu-\mu\in\Delta_+$, and the other half $\nu-\mu\in \Delta_-$.  This is a general feature.  If there is a relevant $\alpha_i$-string in $\Delta_+$, there will be another in $\Delta_-$, with the order of the roots reversed and the roots negated. We will therefore restrict attention to the case $\nu-\mu\in\Delta_+$.  For every nontrivial constraint we find, there will exist another with $\nu-\mu\in\Delta_-$.  Eqn.~(\ref{NonTrivIII}) shows that when $\nu-\mu= \beta\in\Delta_+$, the constraint is $\mu_i \geq d_i[\beta]$; and if $\nu-\mu= -\beta$, the condition becomes $\mu_i \geq h_i[\beta] = d_i[\beta]+\beta_i$.  For completeness, we list both nontrivial constraints in our Table~\ref{nontrivialConditions}. 

We analyze the algebras in alphabetical turn. The resulting list of roots giving rise to nontrivial constraints on $c_{\theta,\mu}^{\nu\not=\mu}$ and ${}^{(k)}N_{\theta,\mu}^{\nu\not=\mu}$ is remarkably short, as highlighted in Table~\ref{nontrivialConditions}. 

\vskip.5cm
\renewcommand{\arraystretch}{1.4}
\begin{table}[!ht]
\begin{center}
    \begin{tabular}{ | l | l | l | l |}
    \hline
  $X_r$ &\ $\beta\ \in\ \Delta_+$ &\ $\nu-\mu=\beta$ &\ $\nu-\mu=-\beta$ \\ \hline\hline
   $A_r$ &\  --- &\ --- \ &\ --- \ \\ \hline
  $B_r$ &\ $\alpha_m + \alpha_{m+1} + \dots + \alpha_r$ \ \ $(1\leq m< r)$&\ $\mu_{r} \geq 1$ &\ $\mu_{r} \geq 1$ \ \\ \hline
    $C_r$ &\ $\alpha_m+ 2\alpha_{m+1} + \dots +2\alpha_{r-1}+ \alpha_r$ \ \ $(1\leq m\leq r-2$ )&\ $\mu_{m} \geq 1$ &\ $\mu_{m} \geq 1$ \  \\ \hline
  $D_r$ &\  --- &\ --- &\ --- \  \ \\ \hline
$E_r$ &\  --- &\ --- &\ --- \ \ \\ \hline
$F_4$ &\ $\alpha_2+\alpha_3$, $\alpha_1+\alpha_2+\alpha_3$, $\alpha_1+2\alpha_2+3\alpha_3 +2\alpha_4$ &\ $\mu_3\geq 1 $ &\ $\mu_3\geq 1 $\ \\ 
\cline{2-4}
  \   &\ $\alpha_2+2\alpha_3+\alpha_4$, $\alpha_1+\alpha_2+2\alpha_3+\alpha_4$,  $\alpha_1+2\alpha_2+2\alpha_3+\alpha_4$ &\  $\mu_4\geq 1 $ &\ $\mu_4\geq 1 $ \  \\ \hline
$G_2$ &\ $\alpha_1+\alpha_2$ &\ $\mu_2\geq 2 $ &\ $\mu_2\geq 1$  \  \\ 
\cline{2-4}
  \   &\ $\alpha_1+2\alpha_2$ &\  $\mu_2\geq 1 $ &\ $\mu_2\geq 2 $  \   \\ \hline
    \end{tabular}
\vskip.5cm\caption{\label{nontrivialConditions} Positive roots $\beta$ and the depth-rule constraints they impose on $c_{\theta,\mu}^{\nu\not=\mu}$ and ${}^{(k)}N_{\theta,\mu}^{\nu\not=\mu}$. The last 2 columns give the constraints $\mu_i \geq d_i[\beta]$ and $\mu_i \geq h_i[\beta] = d_i[\beta]+\beta_i$, imposed when $\nu-\mu=\beta$ and $\nu-\mu= -\beta$, respectively.}
\end{center}  
\end{table}

Simple use of Table~\ref{nontrivialConditions} computes off-diagonal adjoint tensor product coefficients $c_{\theta,\mu}^\nu$ for $\theta, \mu, \nu\in P_+$, and off-diagonal adjoint fusion coefficients ${}^{(k)}N_{\theta,\mu}^\nu$ for $\theta, \mu, \nu\in P_+^k$. In both cases, the coefficient is either 1 or 0.  It is 0 when $\nu-\mu\not\in \Delta$, and is usually 1 when $\nu-\mu\in \Delta$ (e.g., it is always 1 for simply-laced algebras $A_r, D_r, E_{6,7,8}$).  However, when $\nu-\mu=\pm\beta$, with $\beta$ listed in Table~\ref{nontrivialConditions}, the corresponding extra condition must be satisfied for a coefficient 1.  

\vskip.5cm\noindent{$\mathbf{X_r=A_r}$}\\ 
The positive roots of $A_r$ are:
\bens
L_{m,n} :=  \alpha_m+\alpha_{m+1}+\ldots +\alpha_n = -\Lambda^{m-1} +\Lambda^m+\Lambda^n -\Lambda^{n+1} \ ,\ \ (1\leq m\leq n\leq r)\ . 
\eens 
The expansion in terms of simple roots allows us to find the depths $d_i[\beta]$. The expression in terms of fundamental weights gives us the Dynkin labels $\beta_i$.  Comparing the quantities, we can use (\ref{NonTrivII}) to decide if a nontrivial depth-rule constraint $\mu_i\geq d_i[\beta]$ arises. 
 
With $\beta=L_{m,n}$, the only nonzero depths $d_i[\beta]>0$, are $d_{m-1}[\beta] =d_{n+1}[\beta] = 1$. But $\beta_{m-1}=\beta_{n+1} = -1$. 
Since neither situation 1) nor situation 2) of eqn.~(\ref{NonTrivII}) applies, there are no roots producing nontrivial cancellations in the $A_r$ case. 

\vskip.5cm\noindent{$\mathbf{X_r=B_r}$}\\  
The positive roots of $B_r$ are:
\begin{align}
&L_{m} = \sum_{m \leq \ell \leq r} \alpha_\ell = -\Lambda^{m-1}+(1+\delta_{m,r})\Lambda^m, \ \ \ (1 \leq m \leq r);\nn
&M_{m,n} = \sum_{m \leq \ell < n} \alpha_\ell = - \Lambda^{m-1}+\Lambda^m+\Lambda^{n-1} -(1+\delta_{n,r})\Lambda^{n},  \ \ \ (1 \leq m < n \leq r); \nn
&N_{m,n} = \sum_{m\leq \ell < n}\alpha_\ell+ 2 \sum_{n \leq \ell \leq r} \alpha_\ell =   - \Lambda^{m-1}+\Lambda^m+(1+\delta_{n,r})\Lambda^n -\Lambda^{n-1}, \ \ \ (1 \leq m < n \leq r).  \nonumber
\end{align} 

First, consider $\beta=L_m$. We have $d_{m-1}[L_m] = 1$, but $\beta_{m-1}=-1$. However, $\beta_r=2$, while $d_r[L_m]=1$; producing the nontrivial constraint $\mu_r\geq 1$. 

$\beta= M_{m,n}$ has nonzero depths $d_{m-1}[M_{m,n}]= d_n[M_{m,n}] = 1$, and $\beta_{m-1}= -1$ and $\beta_n=-1$. No nontrivial constraints result. 

For $\beta=N_{m,n}$, we have $d_{m-1}[N_{m,n}] = d_{n-1}[N_{m,n}] = 1$, but $\beta_{m-1} = \beta_{n-1} = -1$, so no nontrivial constraints result.

\vskip.5cm\noindent{$\mathbf{X_r=C_r}$}\\   
The positive roots of $C_r$ are:
\begin{align}
&L_{m} = \sum_{m\leq \ell < r} \alpha_\ell + \alpha_r =  -\Lambda^{m-1}+\Lambda^m-\Lambda^{r-1}+\Lambda^r, \ \  (1 \leq m \leq r); \nn
&M_{m,n} = \sum_{m \leq \ell < n} \alpha_\ell =  - \Lambda^{m-1}+\Lambda^m+\Lambda^{n-1} -\Lambda^{n}, \ \ (1 \leq m < n \leq r);\nn
&N_{m,n} = \sum_{m \leq \ell < n}\alpha_\ell+ 2 \sum_{m \leq \ell < r} \alpha_\ell + \alpha_r = - \Lambda^{m-1}+\Lambda^m-\Lambda^{n-1}+\Lambda^{n}, \ \ (1 \leq m < n \leq r). \nonumber
\end{align}

Condition 1) of eqn.~(\ref{NonTrivIII}) is fulfilled when the root is $\beta = \alpha_i + 2 \alpha_{i+1} \dots + 2 \alpha_{r-1}+ \alpha_r = -\Lambda^{i-1} + \Lambda^{i+1}$. Explicitly, we have $\beta_i=0$ and $h_i[\beta] = d_i[\beta] =1$. As before, this implies the nontrivial constraint $\mu_i \geq 1$, with $\beta = \nu - \mu$.  The depth rule conditions are not all trivial for $C_{r}$.  

\vskip.5cm\noindent{$\mathbf{X_r=D_r}$}\\ 
The positive roots of $D_r$ are:
\begin{align}
&L_{m} = \sum_{m \leq \ell \leq r-2} \alpha_\ell + \alpha_r = -\Lambda^{m-1}+\Lambda^m -\Lambda^{r-1}+\Lambda^r, \ \ \ (1 \leq m  < r); \nn
&M_{m,n} = \sum_{m < \ell < n} \alpha_\ell =  - \Lambda^{m}+\Lambda^{m+1}+\Lambda^{n-1} -\Lambda^{n}-\delta_{n,r-1}\Lambda^{r}, \ \ \  (1 \leq m < n \leq r);\nn
&N_{m,n} = \sum_{m\leq \ell < n}\alpha_\ell+  \sum_{n \leq \ell < r-1} 2\alpha_\ell +\alpha_{r-1}+ \alpha_r = - \Lambda^{m-1}+\Lambda^m-\Lambda^{n-1}+\Lambda^{n}, \ \ \ (1 \leq m < n \leq r). \nonumber
\end{align}

The maximum height is $h_i[\beta] =1$, by inspection.   This requires either that $d_{i-1}[\beta] = 1$ and $d_{i+1}[\beta] =0$ or that $d_{i-1}[\beta] = 1$ and $d_{i+1}[\beta] =1$ or  $d_{i-1}[\beta] = 1$ and $d_{i+1}[\beta] =0$. Consideration of each of these scenarios indicates that $\beta_i >0$.  Thus condition 1) of eqn.~(\ref{NonTrivII})  fails. Likewise for condition 2), the depth has a maximum value of $d_{i}[\beta]=1$, by inspection. This implies that   $h_{i-1}[\beta] = 1$ and $h_{i+1}[\beta] =1$; $h_{i-1}[\beta] = 1$ and $h_{i+1}[\beta] =0$; $h_{i-1}[\beta] = 0$ and $h_{i+1}[\beta] =1$; or $h_{i-1}[\beta] = 0$ and $h_{i+1}[\beta] =0$. These scenarios then imply that $\beta_i <0$, so condition 2) of  eqn.~(\ref{NonTrivIII}) also fails. Thus there are no nontrivial constraints for algebras $D_r$.

\vskip.5cm\noindent{$\mathbf{X_r=E_r}$}\\ 
The $E_6$, $E_7$, and $E_8$ algebras can be more easily treated by considering short and long strings of weights. A string of weights consists of those weights related by the addition or subtraction of some $i$-th simple root. For example $\lambda$, $\lambda+\alpha_i$, $\lambda+\alpha_i+\alpha_j$ is not a valid string when $i \not= j$. Strings of two weights are called short strings, with weights denoted $\beta^+$ and $\beta^-$, such that $\beta^+-\beta^- = \alpha_i$. The nontriviality conditions of eqn.~(\ref{NonTrivIII}) will be considered for these weights; it will be demonstrated that weights in short strings for type $E_{6,7,8}$ algebras contribute only trivial conditions. Those strings that contain more than two weights have simple roots on either end, which can be seen by inspection. The nontrivial constraints introduced by these strings will then be associated with the null weight, which only occur in the diagonal tensor or fusion product coefficients.

It is noted that $\beta^+$ has a depth of zero, $d_i[\beta^+]=0$ or $\beta+\alpha_i \not\in \Delta$, and a height of $1$, $h_i[\beta^+]=1$, by definition. If there is a nontrivial constraint then condition 1) of eqn.~(\ref{NonTrivIII}) $\beta_i^+ \leq 0$, must be satisfied. When $\beta_i^+ < 0$ then there must be a label of $\beta^-$ such that $\beta_i^- \leq -3$, but there is no such weight in the adjoint highest weight system. It must be the case that $\beta_i^+ = 0$ indicating that $\beta_i^- = -2$, however these only appear in strings whose highest and lowest weights are simple roots and are contained in strings of 3 weights with the nontrivial condition appearing in the null weight case.

Condition 2) of eqn.~(\ref{NonTrivIII}) can also lead to a nontrivial constraint coming from $\beta^-$. The depth condition $d_i[\beta^-] =1>0$ is satisfied by the definition of $\beta^-$, however imposing $\beta^-_i >0$ requires that there is a label $\beta^+_i>2$ that does not appear. 

Thus for $X_r=E_r$,  the conditions are all trivial.

\vskip.5cm\noindent{$\mathbf{X_r=F_4}$}\\ 
$F_4$ is treated similarly to $E_r$. Nontrivial conditions will be considered for two weights $\beta^-$ and $\beta^+$ being the highest and lowest elements, respectively, of a short string. It will be demonstrated that short strings for $F_4$ imply trivial conditions.  This then leads to the prescription of nontrivial constraints by longer strings of $F_4$ weights.

Condition 2) of eqn.~(\ref{NonTrivIII}) has the depth condition $d_i[\beta^-] =1>0$ satisfied by $\beta^-$,  by definition.   However,  imposing $\beta^-_i >0$ requires that there is a label $\beta^+_i>2$ that does not appear. 

As before $\beta^+$ has a depth of zero, $d_i[\beta^+]=0$, and a height of $1$, $h_i[\beta^+]=1$. If there is a nontrivial constraint then condition 1) of eqn.~(\ref{NonTrivIII}) must be satisfied, that is we require $\beta_i^+ \leq 0$. When $\beta_i^+ < 0$ then there must be a label of $\beta^-$ such that $\beta_i^- \leq -3$, but there is no such weight. Then it must be the case that $\beta_i^+ = 0$ indicating that $\beta_i^- = -2$, these appear in certain cases, which we list in Table~\ref{stringsF4}. Ignored are those cases corresponding to simple roots, since the nontrivial condition is only relevant for diagonal fusion and tensor products. 

We find strings of 3 positive roots $\{\beta-\alpha_i, \beta, \beta+\alpha_i\}$, where $\beta$ is listed in Table~\ref{nontrivialConditions}.  Table~\ref{stringsF4} displays the Dynkin labels, showing the appearance of the values -2 (+2)  for the $i$-th label of the lowest (highest) root in each string.  

\vskip.5cm
\renewcommand{\arraystretch}{1.4}
\begin{table}[!ht]
\begin{center}
    \begin{tabular}{ | l | l | l | l |}
    \hline
 $\beta-\alpha_i$ &\ $\beta$  &\ $\beta+\alpha_i$  &\ $\mu_i\geq d_i[\beta]$  \\ \hline\hline
 (-1,2,-2,0) &\ $\alpha_2+\alpha_3$ &\ (-1,0,2,-2) &\ $\mu_3\geq 1$ \\\hline
  (1,1,-2,0) &\  $\alpha_1+\alpha_2+\alpha_3$ &\ (1,-1,2,-2) &\ $\mu_3\geq 1$ \\ \hline
  (0,1,-2,2) &\  $\alpha_1+2\alpha_2+3\alpha_3 +2\alpha_4$ &\ (0,-1,2,0) &\ $\mu_3\geq 1$\\ \hline
 (-1,0,2,-2) &\ $\alpha_2+2\alpha_3+\alpha_4$  &\  (-1,0,0,2) &\ $\mu_4\geq 1$ \\ \hline
(1,-1,2,-2) &\ $\alpha_1+\alpha_2+2\alpha_3+\alpha_4$   &\ (1,-1,0,2) &\ $\mu_4\geq 1$ \\ \hline
(0,1,0,-2) &\ $\alpha_1+2\alpha_2+2\alpha_3+\alpha_4$ &\ (0,1,-2,2) &\ $\mu_4\geq 1$ \\\hline
    \end{tabular}
\vskip.5cm\caption{\label{stringsF4} $\alpha_i$-strings $\{\beta-\alpha_i, \beta, \beta+\alpha_i\}$ of $F_4$ positive roots, where $\beta$ produces a nontrivial depth-rule constraint (see Table~\ref{nontrivialConditions}). The  strings are listed in the same order as the roots $\beta$ are given in Table~\ref{nontrivialConditions}.  The  Dynkin labels of a root $\gamma$ are provided in the notation $(\gamma_1, \gamma_2, \gamma_3, \gamma_4)$. }
\end{center}  
\end{table}

\vskip1cm
\noindent{\bf Acknowledgments}\hfb
This research was supported in part by a Discovery Grant (MW) from the Natural Sciences and Engineering Research Council (NSERC) of Canada. This paper was completed while MW was visiting the Statistical Physics Group at SISSA; he thanks them for their hospitality.  
\vskip1cm




\vskip1.5cm\noindent\today


\end{document}